# Preprint: Virtual Reality Assistant Technology for Learning Primary Geography

Zhihan Lv[1] and Xiaoming Li[1,2,3]

1. Shenzhen Institutes of Advanced Technology, Chinese Academy of Science, China.
2. Shenzhen Research Center of Digital City Engineering, Shenzhen, China
3. Key Laboratory of Urban Land Resources Monitoring and Simulation, Ministry of Land and Resources, Shenzhen, China
lvzhihan@gmail.com

**Abstract.** This is the preprint version of our paper on ICWL2015. A virtual reality based enhanced technology for learning primary geography is proposed, which synthesizes several latest information technologies including virtual reality(VR), 3D geographical information system(GIS), 3D visualization and multimodal human-computer- interaction (HCI). The main functions of the proposed system are introduced, i.e. Buffer analysis, Overlay analysis, Space convex hull calculation, Space convex decomposition, 3D topology analysis and 3D space intersection detection. The multimodal technologies are employed in the system to enhance the immersive perception of the users.

## 1 Introduction

With the developments in web, mobile and virtual reality technologies as well as mass adoption of smart and mobile devices by all people in the current society, significant opportunities have emerged for e-learning applications [16]. Location-based mobile phone software has been proved to have the potential in support-ing biology, geography and math lessons for students in schools [12]. The latest technologies from video game industry make it becomes easy and efficient to integrate the state-of-the-art graphics/interface and the educational aspect with poor graphics and interaction [1] [25]. The technologies of video games have attracted considerable attention to enhance user participation and motivation which should be a key concern in the design of an interactive human-computer-interaction(HCI) interface [5] of the learning enhanced system [10]. Technology enhanced learning brings multidisciplinary, interdisciplinary, and pandisciplinary educational content in different forms [2] oriented to different culture [40]. The dynamic learning environments can deliver educational benefits as educational offerings based upon various characteristics of individual learners [27]. Synthe-sizing multimedia operations and knowledge sharing aspects has also attracted attention from the research community [15]. The combinations of technology enhanced learning and other applied information field have been considered as promising practical research topics [36]. It has been argued that integration of concepts from GIS into Information Technology (IT) can provide impressive opportunities for education [38]. The guidelines for the design of 3D virtual learning spaces has been suggested before [26]. The related evaluation process has been proposed in previous research [16] [6]. Besides, a lot of other previous related researches have also inspired our research [28] [35] [39] [41] [30] [13] [9] [31] [29] [52].

In virtual reality technology trend, geospatial data visualization has never been developing rapidly [34]. With the development of VR (Virtual Reality) technology and widely applications in various areas, the requirements to VR are also increasing rapidly. Virtual Reality Geographical Information System (VRGIS), a combination of geographic information system and virtual reality technology [14] has become a hot topic. Accordingly, '3-D modes' has been proved as a faster decision making tool with fewer errors [33]. Meanwhile, new user interfaces for geo-databases is also expected [3]. With the popularity of network, the VRGIS platform based on the network environment also becomes a trend. The application of VRML, X3D and other online VR technologies have achieved networking of VR systems [37] [23] [21] [20].

WebVRGIS is preferred in practical applications, especially by the geography and urban planning [22] [19], which is based on WebVR [24]. Urban simulation is becoming widely noticed nowadays, and some simulation systems have been developed in this area, e.g. ArcView3D Analyst, Imagine Virtual GIS, GeoMedia, etc. WebVRGIS engine supports steadily real time navigation in virtual scenes which are constructed with massive, multi-dimensional data from various sources [55] [18] [47] [54] [46] [17] [48] [44] [45] [32] [53] [49] [50] [42] [43] [51]. 3D urban landscape database with various data sources can be produced to implement spatial analysis and 3D visualization and published in the Internet environment [11] [7] [8]. In this research, we plan to use virtual reality geographic information system technology as an education tool to teach the geography courses.

## 2  System Overview

3D geometrical analysis education model includes 8 sub-modules: entity set operation, buffer analysis, overlay analysis, space convex hull calculation, space convex decomposition, 3D topology analysis, 3D Minkowski sum calculation, and 3D space intersection detection, as shown in figure 1.

## 3  Functions

Entity set operation includes 4 sub-modules: union, meet, difference, and equilibrium difference. In the space entity, union outputs a new entity including all input entities. Meet outputs the common part of all input entities. Difference outputs spatial features included in the former input entity but excluded from the next space entity.

Buffer analysis includes point, line, face, and body. Buffer analysis is served as one of spatial analysis tools used to solve proximity-related problems. It is used to realize spatial extension of a spatial entity, determine the scope of influence of this entity in 3D space, and reflect the gradual change law of its influence with distance variation. For instance, the buffer of special sphere with the radius as r (buffer distance R) is a ball with the radius as r+R.

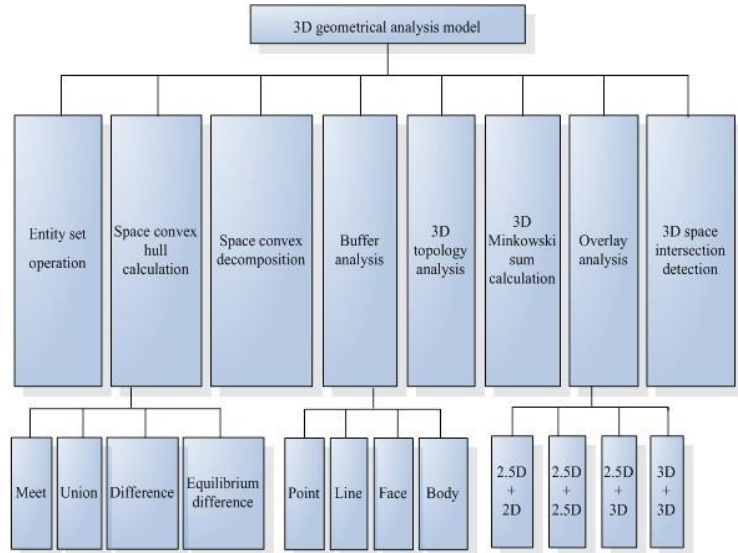

**Fig. 1.** 3D Geometrical Analysis Model

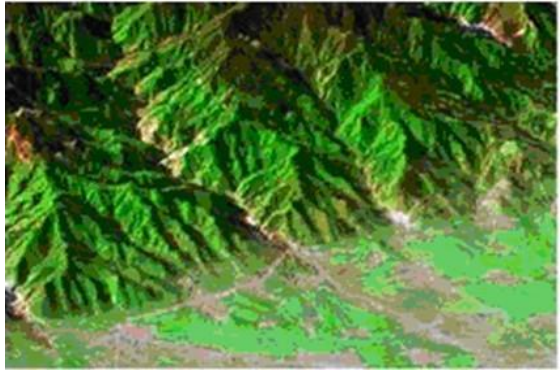

**Fig. 2.** Overlay Analysis

Overlay analysis refers to an analytical method in which two or more geographic element photos in the same region are overlapped under a unified spatial reference system to generate multiple attributive characters of this region. Overlay analysis can be used for new classification of multiple attributes generated after such overlay, so as to establish spatial correspondence among geological objects, and extract quantitative characteristics of some topics within the scope of a certain region.

Space convex hull calculation: in 3D figure application, we often encounter such an issue: determine a convex polyhedron with the minimum volume and including several given points in the space. It is also a form of the minimum coverage problems to calculate the algorithm of convex hull including spatial point sets. It is widely applied in the modeling of buildings.

Space convex decomposition: convex decomposition of a concave polygon is one of basic problems about computation geometry. It is applied in many fields. It realizes the function of decomposing complicated space convex into simple convex polygon. Most existing algorithms are of overall decomposition algorithm. However, fewer researches are made on partial decomposition algorithm. The overall decomposition algorithm is time-consuming. Therefore, it cannot meet application needs of all projects.

3D topology analysis: space topological relation is an important part of spatial analysis. How to effectively generate and judge space topological relation has direct impact on the efficiency of spatial information system. The establishment of 3D topological relation makes it easy to realize various space operations and information inquiry. Complicated surface features can be described with various bodies filled in the space, hook faces constituting those bodies, boundary rings constituting those hook faces, arcs constituting those rings, and nodes on those arcs. Generally speaking, body is basic constitution for the target entity. Any complicated entities are deemed as composed of bodies (natural or artificial). Body, face, line and point are dynamic concepts. They are interconvertible under different scales or different research emphases. The following 6 groups of relations are used to describe the space topological relation of 3D vector structure in the spatial information system: relation between complicated surface feature and body, relation among body, complicated surface feature, and hook face, relation among hook face, ring, and body, relation among ring, arc and hook face, relation among arc, node, and ring, and relation between node and arc.

3D space intersection detection: while simulating 3D animation scenes, numerous collision phenomena will be encountered. Essentially, these collision phenomena are points, lines, faces, and bodies. It is only necessary to judge whether they intersect in 3D space, and work out intersecting points, intersecting lines, intersecting faces, and intersecting bodies.

## 4 Application Scenarios

The oriented city region simulation WebVRGIS engine is developed based on OpenGL and C++, which integrates VR and GIS seamlessly and supports massive data.

As shown in figure 4, the user is watching the virtual geographical scene through the HMD. The HMD is the VR glasses shell by which users could watch the anaglyph 3D scene generated from smartphone screen. The convergence-to-face part of the HMD for light blocking is made by soft holster filling of sponge, so it's not oppressive at all. In addition, it can modify the pupil distance (PD) and depth of field (DOF), so it is suitable for the users with different myopia degree and PD. The input methods include head motion and remote controller. The head motion on VR glasses only supports the rotations around three axis which are controlled by gyroscope sensor of the smartphone. The head rotation actions are synchronous with the rotation of the camera view in the VR scene, which brings the immersive perception to the user in real time. Meanwhile, the remote controller is used to input the displacement of the 3D scene as well as manipulate the menu of the software configuration.

## 5 Conclusion

WebVRGIS is used to show the geographical information intuitively as an enhanced learning tool. Geographical information has several characteristics, ie. large scale, diverse predictable and real-time, which falls in the range of definition of Big Data defined by Faye Briggs [4]. Virtual reality is a promising and suitable technology to represent geographical bigdata. In future, the new technology will enhance the learning process, for example, multiple users may support the teacher to train a class at the same time.

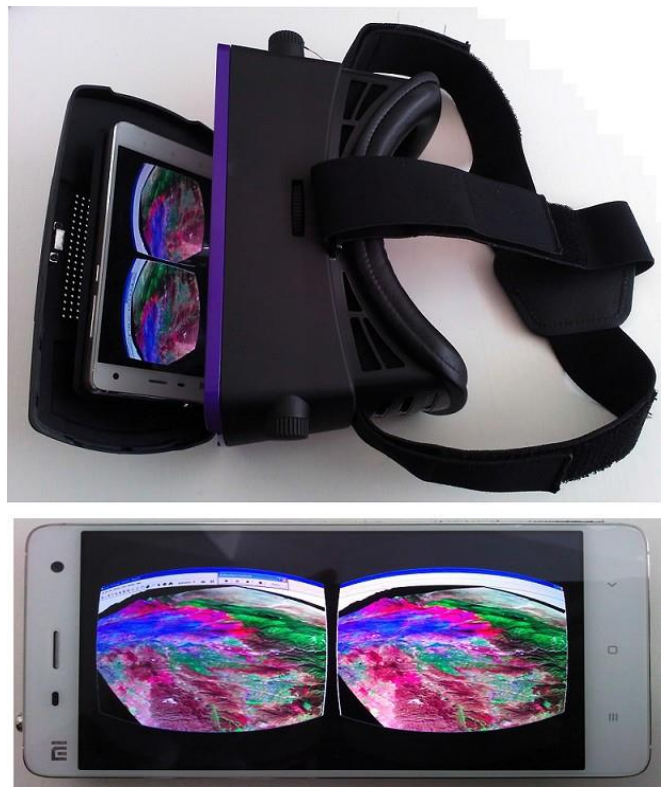

**Fig. 3.** VRGIS Running on Virtual Reality Glassess Device

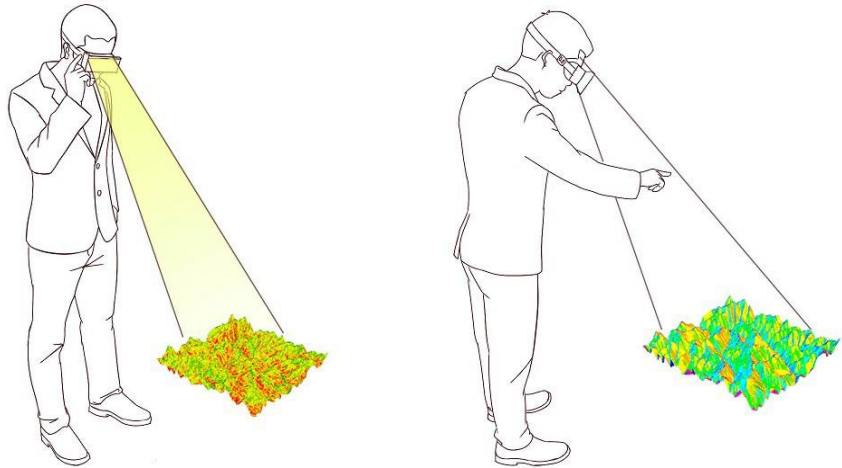

**Fig. 4.** Left: the user wears the virtual reality Glasses to look at the virtual 3D GIS; Right: the user wears the virtual reality Glasses and use touch-less interaction to manipulate the 3D GIS